\documentclass[conference,compsoc]{IEEEtran}

\usepackage[T1]{fontenc}
\usepackage{amsmath,amssymb}
\usepackage{booktabs}
\usepackage{array}
\usepackage{graphicx}
\usepackage{xspace}
\usepackage{url}
\usepackage{pgfplots}
\pgfplotsset{compat=1.18}
\usetikzlibrary{positioning,calc,arrows.meta,matrix,shapes.geometric,fit,backgrounds}
\definecolor{psBlue}{HTML}{2A78D6}
\definecolor{psOrange}{HTML}{EB6834}
\definecolor{psGray}{HTML}{8A8A86}
\definecolor{psInk}{HTML}{52514E}
\pgfplotsset{
  policysniff/.style={
    width=0.95\columnwidth, height=5.2cm,
    ybar, bar width=9pt, enlarge x limits=0.25,
    ymin=0, ymax=100, ytick={0,25,50,75,100},
    enlarge y limits={upper, value=0.06},
    axis lines*=left, axis line style={psGray, line width=0.5pt},
    tick align=outside, tick pos=left, major tick length=2pt, tick style={psGray, line width=0.5pt},
    ymajorgrids, grid style={psGray!20, line width=0.3pt},
    ylabel={balanced accuracy (\%)}, ylabel style={font=\scriptsize, color=psInk},
    tick label style={font=\scriptsize, color=psInk},
    legend style={font=\scriptsize, draw=none, legend columns=-1, /tikz/every even column/.append style={column sep=5pt}},
    legend cell align=left,
    error bars/error mark options={rotate=90, mark size=1.5pt, psGray, line width=0.5pt},
  },
  psbar/.style={draw=none},
  psaxisplain/.style={
    axis lines*=left, axis line style={psGray, line width=0.5pt},
    tick align=outside, tick pos=left, major tick length=2pt, tick style={psGray, line width=0.5pt},
    ymajorgrids, grid style={psGray!20, line width=0.3pt},
    tick label style={font=\scriptsize, color=psInk},
    xlabel style={font=\scriptsize, color=psInk}, ylabel style={font=\scriptsize, color=psInk},
    legend style={font=\scriptsize, draw=none, /tikz/every even column/.append style={column sep=5pt}},
    legend cell align=left,
  },
}
\newcommand{\pslegendbelow}[1]{\node[anchor=north, inner sep=0pt] at ([yshift=-3pt]current axis.outer south) {\pgfplotslegendfromname{#1}};}
\tikzset{
  >={Latex[length=1.5mm,width=1.2mm]},
  psline/.style={line width=0.5pt},
  psnode/.style={psline, draw=psGray!65, fill=psGray!8, rounded corners=2pt, align=center, inner sep=3pt, minimum height=6mm, font=\scriptsize, text=black},
  psblue/.style={draw=psBlue!60, fill=psBlue!8},
  psorange/.style={draw=psOrange!60, fill=psOrange!8},
  pspanel/.style={psline, draw=psGray!55, fill=psGray!5, rounded corners=3pt, inner sep=4pt},
  pspanelblue/.style={pspanel, draw=psBlue!35, fill=psBlue!5},
  pspanelorange/.style={pspanel, draw=psOrange!35, fill=psOrange!5},
  pstitle/.style={font=\footnotesize, text=black, anchor=west},
  pslabel/.style={font=\scriptsize, text=psInk},
  psglyph/.style={font=\tiny, text=psInk},
  psflow/.style={->, psline, draw=psGray!85, shorten >=1pt},
  psflowdashed/.style={psflow, dashed},
  pslink/.style={psline, draw=psGray!70, dashed},
  psbox/.style={psnode},
  pslab/.style={pslabel, anchor=east},
}

\usepackage[caption=false,font=footnotesize,labelfont=sf,textfont=sf]{subfig}
\usepackage[hidelinks]{hyperref}
\usepackage{cite}

\newcommand{\AmBA}{86.11\%\xspace}
\newcommand{\AmBASwap}{86.11\%\xspace}
\newcommand{\AmBits}{25{,}920/25{,}920\xspace}
\newcommand{\AmBitsSwap}{25{,}920/25{,}920\xspace}
\newcommand{\AmCI}{[75.00\%, 97.22\%]\xspace}
\newcommand{\AmFamBuild}{83.33\%\xspace}
\newcommand{\AmFamFix}{100.00\%\xspace}
\newcommand{\AmFamLease}{75.00\%\xspace}

\newcommand{\AmPerm}{44.44\%\xspace}
\newcommand{\AmPermSwap}{44.44\%\xspace}
\newcommand{\AmSham}{53.12\%\xspace}
\newcommand{\BhAllBA}{91.08\%\xspace}
\newcommand{\BhAllCount}{1093/1200\xspace}
\newcommand{\BhDecBA}{88.82\%\xspace}
\newcommand{\BhDecClusters}{79\xspace}
\newcommand{\BhDecCondApproved}{30\xspace}

\newcommand{\BhDecCondReadOnly}{64\xspace}
\newcommand{\BhDecCondRevoked}{65\xspace}
\newcommand{\BhDecCondUnapproved}{11\xspace}
\newcommand{\BhDecCount}{302/340\xspace}
\newcommand{\BhDecRestBA}{91.98\%\xspace}

\newcommand{\BhDecTasks}{170\xspace}
\newcommand{\BhExecBA}{90.00\%\xspace}
\newcommand{\BhExecCount}{54/60\xspace}
\newcommand{\BhExecTasks}{30\xspace}

\newcommand{\BhFullAllClusters}{24\xspace}

\newcommand{\BhFullAllTasks}{120\xspace}
\newcommand{\BhFullBA}{80.30\%\xspace}
\newcommand{\BhFullClusters}{15\xspace}
\newcommand{\BhFullCount}{53/66\xspace}
\newcommand{\BhFullRestBA}{86.78\%\xspace}

\newcommand{\BhFullTasks}{33\xspace}
\newcommand{\BhFullTokens}{3\xspace}
\newcommand{\BhOrderCorr}{$0.003$\xspace}
\newcommand{\BlBuildConfirm}{0.83\%\xspace}
\newcommand{\BlBuildPlacebo}{45.83\%\xspace}
\newcommand{\BlFixConfirm}{16.67\%\xspace}
\newcommand{\BlFixPlacebo}{94.17\%\xspace}
\newcommand{\BlLeaseConfirm}{0.00\%\xspace}
\newcommand{\BlLeasePlacebo}{43.33\%\xspace}

\newcommand{\BrDecodeSteps}{360\xspace}

\newcommand{\BrMoeKernels}{216\xspace}
\newcommand{\BsBenignConfirm}{43.06\%\xspace}
\newcommand{\BsBenignDrop}{9.72\xspace}
\newcommand{\BsBenignPlacebo}{52.78\%\xspace}
\newcommand{\BsBuildAuth}{79.17\% (19/24)\xspace}
\newcommand{\BsBuildConfirm}{0.00\% (0/24)\xspace}
\newcommand{\BsBuildDrop}{66.67\xspace}
\newcommand{\BsBuildPlacebo}{66.67\% (16/24)\xspace}
\newcommand{\BsFixAuth}{95.83\% (23/24)\xspace}
\newcommand{\BsFixConfirm}{8.33\% (2/24)\xspace}
\newcommand{\BsFixDrop}{62.50\xspace}
\newcommand{\BsFixPlacebo}{70.83\% (17/24)\xspace}
\newcommand{\BsLeaseAuth}{87.50\% (21/24)\xspace}
\newcommand{\BsLeaseConfirm}{0.00\% (0/24)\xspace}
\newcommand{\BsLeaseDrop}{75.00\xspace}
\newcommand{\BsLeasePlacebo}{75.00\% (18/24)\xspace}

\newcommand{\CgSevenECorr}{0.8832\xspace}
\newcommand{\CgSevenEFive}{75.00\%\xspace}
\newcommand{\CgSevenEFiveCI}{[61.11\%, 88.89\%]\xspace}
\newcommand{\CgSevenEFiveCount}{27/36\xspace}

\newcommand{\CgSevenESingle}{72.22\%\xspace}

\newcommand{\CgSixCorr}{0.8824\xspace}
\newcommand{\CgSixFive}{100.00\%\xspace}
\newcommand{\CgSixFiveCI}{[100.00\%, 100.00\%]\xspace}
\newcommand{\CgSixFiveCount}{36/36\xspace}

\newcommand{\CgSixOne}{81.67\%\xspace}
\newcommand{\CpNatFivePool}{87.50\%\xspace}
\newcommand{\CpNatOnePool}{76.94\%\xspace}
\newcommand{\CpPadFivePool}{52.78\%\xspace}
\newcommand{\CpPadOnePool}{50.83\%\xspace}
\newcommand{\CpUniFivePool}{52.78\%\xspace}
\newcommand{\CpUniOnePool}{51.39\%\xspace}
\newcommand{\CtPadFiveGSevenE}{52.78\%\xspace}
\newcommand{\CtPadFiveGSix}{52.78\%\xspace}
\newcommand{\CtPadOneGSevenE}{52.78\%\xspace}
\newcommand{\CtPadOneGSix}{48.89\%\xspace}
\newcommand{\CtUniFiveGSevenE}{55.56\%\xspace}
\newcommand{\CtUniFiveGSix}{50.00\%\xspace}
\newcommand{\CtUniOneGSevenE}{55.00\%\xspace}
\newcommand{\CtUniOneGSix}{47.78\%\xspace}

\newcommand{\LgFourBA}{97.78\%\xspace}

\newcommand{\LgSevenBA}{97.08\%\xspace}
\newcommand{\LgSevenCI}{[93.89\%, 99.03\%]\xspace}
\newcommand{\LgSixBA}{98.33\%\xspace}
\newcommand{\LgSixCI}{[96.81\%, 99.58\%]\xspace}
\newcommand{\LgSixFamBuild}{99.58\%\xspace}
\newcommand{\LgSixFamFix}{100.00\%\xspace}
\newcommand{\LgSixFamLease}{95.42\%\xspace}
\newcommand{\LgSixLen}{59.03\%\xspace}
\newcommand{\LgSixPerm}{50.28\%\xspace}
\newcommand{\LgThreeBA}{97.92\%\xspace}
\newcommand{\PadCyclesSevenE}{107.72\%\xspace}
\newcommand{\PadCyclesSix}{107.87\%\xspace}
\newcommand{\PadEnergySevenE}{69.80\%\xspace}
\newcommand{\PadEnergySix}{68.72\%\xspace}
\newcommand{\PadLatency}{27{,}072\xspace}
\newcommand{\PcBasicBA}{94.17\%\xspace}
\newcommand{\PcBasicCI}{[90.42\%, 97.50\%]\xspace}

\newcommand{\PcBasicCoverage}{100.00\%\xspace}

\newcommand{\PcCmpCostContinuousBA}{95.83\%\xspace}

\newcommand{\PcCmpCostContinuousCoverage}{100.00\%\xspace}

\newcommand{\PcCmpCostContinuousMedianMs}{11846.2\xspace}

\newcommand{\PcCmpCostFiveMsBA}{62.50\%\xspace}

\newcommand{\PcCmpCostFiveMsCoverage}{95.83\%\xspace}

\newcommand{\PcCmpCostFiveMsMedianMs}{3789.9\xspace}

\newcommand{\PcCmpCostNoProbeCoverage}{0.00\%\xspace}

\newcommand{\PcCmpCostNoProbeMedianMs}{2504.1\xspace}

\newcommand{\PcCmpCostOneMsBA}{93.75\%\xspace}

\newcommand{\PcCmpCostOneMsCoverage}{100.00\%\xspace}

\newcommand{\PcCmpCostOneMsMedianMs}{7465.5\xspace}

\newcommand{\PcCompleteBA}{91.67\%\xspace}
\newcommand{\PcCompleteCI}{[83.33\%, 97.92\%]\xspace}

\newcommand{\PcCompleteCoverage}{100.00\%\xspace}

\newcommand{\PcModels}{20\xspace}

\newcommand{\PcQuantization}{NF4\xspace}

\newcommand{\PcTorch}{2.8.0+cu126\xspace}

\newcommand{\PcTransformers}{4.55.4\xspace}

\newcommand{\PgApprovedBaseline}{95.83\%\xspace}
\newcommand{\PgApprovedGuard}{41.67\%\xspace}
\newcommand{\PgApprovedIdentical}{12/12\xspace}
\newcommand{\PgBaselineBA}{95.83\%\xspace}
\newcommand{\PgBaselineBehavior}{48/48\xspace}

\newcommand{\PgBaselineCoverage}{100.00\%\xspace}

\newcommand{\PgDiffCI}{[50.00, 66.67]\xspace}
\newcommand{\PgDiffPP}{58.33\xspace}

\newcommand{\PgGuardAllBA}{37.50\%\xspace}
\newcommand{\PgGuardBehavior}{48/48\xspace}

\newcommand{\PgGuardCondBA}{47.37\%\xspace}
\newcommand{\PgGuardCoverage}{79.17\%\xspace}

\newcommand{\QmActionBA}{97.92\%\xspace}
\newcommand{\QmActionCount}{94/96\xspace}
\newcommand{\QmAllParse}{120/120\xspace}

\newcommand{\QmBlindBA}{97.92\%\xspace}
\newcommand{\QmBlindCI}{[93.75\%, 100.00\%]\xspace}
\newcommand{\QmBlindCount}{47/48\xspace}
\newcommand{\QmBootstrapFloor}{60.00\%\xspace}
\newcommand{\QmCompleteBA}{89.58\%\xspace}
\newcommand{\QmCompleteCI}{[81.25\%, 97.92\%]\xspace}
\newcommand{\QmCompleteCount}{43/48\xspace}
\newcommand{\QmContinuousBA}{97.92\%\xspace}
\newcommand{\QmContinuousS}{10.524\xspace}
\newcommand{\QmContinuousSlow}{4.71$\times$\xspace}

\newcommand{\QmGateFloor}{75.00\%\xspace}
\newcommand{\QmNoProbeBA}{51.04\%\xspace}
\newcommand{\QmNoProbeS}{2.237\xspace}
\newcommand{\QmQuantizedLayers}{18{,}672\xspace}
\newcommand{\QmRequestsPhase}{960\xspace}

\newcommand{\QmSparseFiveBA}{58.33\%\xspace}
\newcommand{\QmSparseFiveS}{3.403\xspace}
\newcommand{\QmSparseFiveSlow}{1.52$\times$\xspace}
\newcommand{\QmSparseOneBA}{94.79\%\xspace}
\newcommand{\QmSparseOneS}{6.627\xspace}
\newcommand{\QmSparseOneSlow}{2.96$\times$\xspace}

\newcommand{\QrBA}{95.83\%\xspace}
\newcommand{\QrCI}{[89.58\%, 100.00\%]\xspace}
\newcommand{\QrCount}{46/48\xspace}
\newcommand{\QrCtrlBA}{43.75\%\xspace}
\newcommand{\QrCtrlCount}{21/48\xspace}
\newcommand{\QrFamFs}{100.00\% (16/16)\xspace}
\newcommand{\QrFamOb}{87.50\% (14/16)\xspace}
\newcommand{\QrFamPr}{100.00\% (16/16)\xspace}
\newcommand{\QrGapAgree}{24/24\xspace}
\newcommand{\QrGapCorr}{0.79\xspace}
\newcommand{\QrGapCtrl}{0.000032\xspace}
\newcommand{\QrGapRef}{0.138624\xspace}
\newcommand{\QrGapTgt}{0.125312\xspace}

\newcommand{\QrOrderCorr}{$-0.055$\xspace}
\newcommand{\QrOrderDetrMs}{80.9\xspace}
\newcommand{\QrOrderRawMs}{82.6\xspace}
\newcommand{\QrPairCorr}{0.78\xspace}
\newcommand{\QrPairSign}{24/24\xspace}

\newcommand{\RlBitAcc}{99.994\%\xspace}
\newcommand{\RlBits}{720-bit\xspace}

\newcommand{\RlPermPolicy}{50.00\%\xspace}
\newcommand{\RlPolicy}{80.56\%\xspace}

\newcommand{\UaGEightDetected}{719\xspace}

\newcommand{\UaGEightFiveBA}{80.56\%\xspace}

\newcommand{\UaGEightFiveCount}{29/36\xspace}

\newcommand{\UgThreeBA}{81.11\%\xspace}
\newcommand{\UgThreeCI}{[71.11\%, 91.67\%]\xspace}
\newcommand{\UgThreeDisjoint}{46.11\%\xspace}
\newcommand{\UgThreeFamBuild}{68.33\%\xspace}
\newcommand{\UgThreeFamFix}{98.33\%\xspace}
\newcommand{\UgThreeFamLease}{76.67\%\xspace}
\newcommand{\UgThreePerm}{50.82\%\xspace}
\newcommand{\UmFourBA}{77.22\%\xspace}
\newcommand{\UmFourCI}{[69.44\%, 85.56\%]\xspace}
\newcommand{\UmFourDisjoint}{56.67\%\xspace}
\newcommand{\UmFourDisjointCI}{[52.22\%, 60.56\%]\xspace}
\newcommand{\UmFourGap}{20.56\xspace}
\newcommand{\UmFourPerm}{50.89\%\xspace}
\newcommand{\UmTwoBA}{80.56\%\xspace}
\newcommand{\UmTwoCI}{[70.00\%, 91.67\%]\xspace}
\newcommand{\UmTwoDisjoint}{53.89\%\xspace}
\newcommand{\UmTwoSource}{77.78\%\xspace}
\newcommand{\UmTwoUniform}{47.78\%\xspace}

\hypersetup{pdftitle={PolicySniff: Inferring Agent Policy States through Shared-GPU Execution},
pdfauthor={Weimin Fu, Minghao Shao, Zeng Wang, Qiying Li, Yuntao Liu, Johann Knechtel, Ozgur Sinanoglu, Muhammad Shafique, Ramesh Karri, Xiaolong Guo}}

\begin{document}

\title{PolicySniff: Inferring Agent Policy States through Shared-GPU Execution}

\author{\IEEEauthorblockN{Weimin Fu\IEEEauthorrefmark{1}, Minghao Shao\IEEEauthorrefmark{2}, Zeng Wang\IEEEauthorrefmark{2}, Qiying Li\IEEEauthorrefmark{1}, Yuntao Liu\IEEEauthorrefmark{1}\\
Johann Knechtel\IEEEauthorrefmark{3}, Ozgur Sinanoglu\IEEEauthorrefmark{3}, Muhammad Shafique\IEEEauthorrefmark{3}, Ramesh Karri\IEEEauthorrefmark{2}, Xiaolong Guo\IEEEauthorrefmark{1}}
\IEEEauthorblockA{\IEEEauthorrefmark{1}Lehigh University\\\IEEEauthorrefmark{2}New York University\\\IEEEauthorrefmark{3}New York University Abu Dhabi}}

\maketitle

\begin{abstract}
LLM agents apply approval settings, such as whether to ask before a consequential tool call, through instructions in their system prompt, and operators keep these settings confidential by restricting access to prompts and responses. Audits check whether an agent obeys its setting, not whether a neighboring process can infer which setting is deployed. PolicySniff infers an agent's approval-policy state through shared-GPU execution. A co-resident attacker that knows the candidate settings and the task first profiles both settings, then times only its own GPU kernels while the agent processes the task. The policy prefix is cached before capture, so the attacker observes processing of a task suffix that is identical under both settings. On an eager sparse mixture-of-experts runtime, where each selected expert runs as separate work, a frozen and blinded study identified the deployed setting at \PcBasicBA\ balanced accuracy on its primary estimand. Across all authorization conditions, accuracy was \BhDecBA\ on tasks for which both settings produce the same decision and \BhDecRestBA\ on tasks where they differ. Hardware replays of expert routing show that its temporal density and layer-and-bucket alignment carry the policy information. Behavioral compliance does not imply configuration confidentiality: deployments that keep their approval setting private should measure its distinguishability on the serving runtime and enforce authorization outside the model.
\end{abstract}

\section{Introduction}\label{sec:intro}

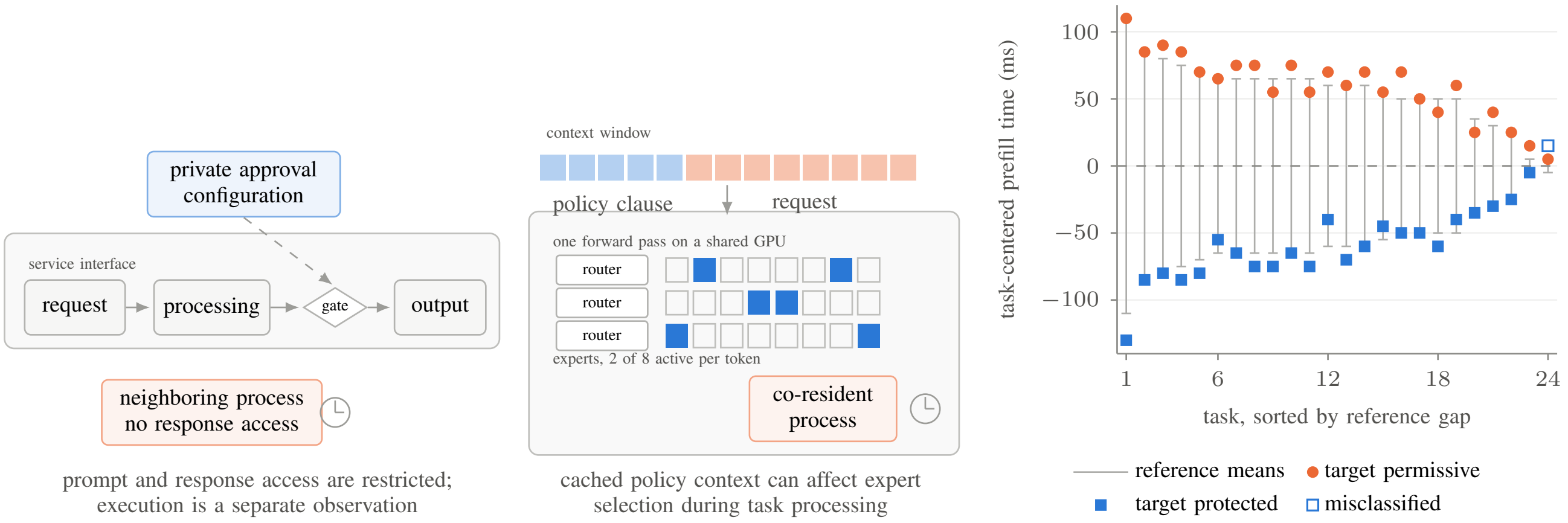
\begin{figure*}[t]
\centering
\subfloat[service access restrictions\label{fig:hero-a}]{%
\begin{tikzpicture}[font=\scriptsize]
\node[psnode, psblue, text width=19mm] (pol) at (2.6,2.6) {private approval\\configuration};
\node[psnode, minimum width=11mm] (req) at (0.75,1.25) {request};
\node[psnode, minimum width=12mm] (proc) at (2.25,1.25) {processing};
\node[psline, draw=psGray!65, diamond, aspect=1.6, inner sep=1pt, fill=white, font=\tiny] (gate) at (3.6,1.25) {gate};
\node[psnode, minimum width=10mm] (out) at (4.75,1.25) {output};
\node[psglyph, anchor=north west] (plabel) at (0.12,1.92) {service interface};
\begin{scope}[on background layer]
\node[pspanel, fit={(req)(proc)(gate)(out)(plabel)}] (cbox) {};
\end{scope}
\node[psnode, psorange, text width=22mm, minimum height=5mm] (obs) at (2.25,0.1) {neighboring process\\no response access};
\draw[psline, draw=psGray!85] (3.6,0.1) circle (0.16) (3.6,0.1)--(3.6,0.23) (3.6,0.1)--(3.7,0.1);
\draw[psflow] (req)--(proc); \draw[psflow] (proc)--(gate); \draw[psflow] (gate)--(out);
\draw[psflowdashed] (pol.south)--(gate.north);
\node[pslabel, anchor=north, text width=46mm, align=center] at (2.75,-0.42) {prompt and response access are restricted; execution is a separate observation};
\end{tikzpicture}}
\hfil
\subfloat[policy in the context window\label{fig:hero-b}]{%
\begin{tikzpicture}[font=\scriptsize]
\node[psglyph, anchor=south west] at (0.05,2.74) {context window};
\foreach \i in {0,...,4} {\fill[psBlue!35] (0.1+\i*0.32,2.4) rectangle ++(0.28,0.28);}
\foreach \i in {5,...,12} {\fill[psOrange!40] (0.1+\i*0.32,2.4) rectangle ++(0.28,0.28);}
\node[pslabel, anchor=north] at (0.9,2.36) {policy clause};
\node[pslabel, anchor=north] at (3.0,2.36) {request};
\foreach \l/\a/\b in {0/1/6, 1/3/4, 2/0/7} {
  \node[psline, draw=psGray!65, rounded corners=1pt, inner sep=1pt, minimum width=10mm, minimum height=3.2mm, fill=white, font=\tiny] (r\l) at (0.78,1.42-\l*0.36) {router};
  \foreach \e in {0,...,7} {
    \ifnum\e=\a \fill[psBlue] (1.48+\e*0.3,1.29-\l*0.36) rectangle ++(0.24,0.26);
    \else\ifnum\e=\b \fill[psBlue] (1.48+\e*0.3,1.29-\l*0.36) rectangle ++(0.24,0.26);
    \else \draw[psline, draw=psGray!55] (1.48+\e*0.3,1.29-\l*0.36) rectangle ++(0.24,0.26);\fi\fi
  }
}
\node[psglyph, anchor=north west] (elabel) at (0.12,0.64) {experts, 2 of 8 active per token};
\node[psglyph, anchor=north west] (glabel) at (0.12,1.92) {one forward pass on a shared GPU};
\node[psnode, psorange, text width=14mm, minimum height=5mm] (spy) at (3.2,-0.1) {co-resident\\process};
\draw[psline, draw=psGray!85] (4.32,-0.1) circle (0.16) (4.32,-0.1)--(4.32,0.03) (4.32,-0.1)--(4.42,-0.1);
\begin{scope}[on background layer]
\node[pspanel, fit={(r0)(r2)(glabel)(elabel)(spy)(4.55,-0.1)}] (gbox) {};
\end{scope}
\draw[psflow] (2.15,2.35)--(2.15,1.98);
\node[pslabel, anchor=north, text width=44mm, align=center] at (2.3,-0.66) {cached policy context can affect expert selection during task processing};
\end{tikzpicture}}
\hfil
\subfloat[what the neighbor measures\label{fig:hero-c}]{%
\begin{tikzpicture}
\begin{axis}[psaxisplain, width=0.36\textwidth, height=5.4cm, unbounded coords=discard,
  xlabel={task, sorted by reference gap}, ylabel={task-centered prefill time (ms)},
  xmin=0.5, xmax=24.5, xtick={1,6,12,18,24},
  ymin=-140, ymax=120, ytick={-100,-50,0,50,100}, axis line style={-},
  legend to name=psleg-hero, legend columns=2]
\addlegendimage{line legend, psGray!70, line width=0.5pt}
\addplot[forget plot, ycomb, psGray!70, line width=0.5pt, mark=-, mark size=1.5pt, mark options={psGray!70, line width=0.5pt}] table[x=x, y=dev_ref_perm] {fig/hero_native.dat};
\addplot[forget plot, ycomb, psGray!70, line width=0.5pt, mark=-, mark size=1.5pt, mark options={psGray!70, line width=0.5pt}] table[x=x, y=dev_ref_prot] {fig/hero_native.dat};
\addplot[only marks, mark=*, mark size=1.6pt, psOrange] table[x=x, y=hit_perm] {fig/hero_native.dat};
\addplot[only marks, mark=square*, mark size=1.6pt, psBlue] table[x=x, y=hit_prot] {fig/hero_native.dat};
\addplot[forget plot, only marks, mark=o, mark size=1.7pt, psOrange, line width=0.6pt, mark options={fill=white}] table[x=x, y=miss_perm] {fig/hero_native.dat};
\addplot[only marks, mark=square, mark size=1.7pt, psBlue, line width=0.6pt, mark options={fill=white}] table[x=x, y=miss_prot] {fig/hero_native.dat};
\legend{reference means, target permissive, target protected, misclassified}
\draw[dashed, psGray, line width=0.5pt] (axis cs:0.5,0) -- (axis cs:24.5,0);
\end{axis}
\pslegendbelow{psleg-hero}
\end{tikzpicture}}
\caption{Policy-state inference outside the response interface. (a) Service access restrictions withhold the configuration and action response from the neighbor. (b) Task execution remains conditioned on the cached policy context; the neighbor times its own kernels. Expert counts are schematic. (c) Blinded NF4 campaign in the agent's GPU instance: each column is a task, gray segments span its two reference means, and filled circles and squares show the permissive and protected target groups; open markers denote misclassified groups, and values are offset by the task midpoint. The decoder classified \QmBlindCount\ groups correctly.}
\label{fig:hero}
\end{figure*}

LLM agents use tools to act on their users' behalf, including sending messages, modifying files, and executing code \cite{DBLP:conf/nips/DebenedettiZBB024}. Agent frameworks govern access to these tools and resources, while the model selects actions in pursuit of a task. Permission to reach a resource leaves further authorization questions: which data may be sent, to whom, and for what purpose. A documented agent incident illustrates this gap: a domain allowlist admitted a request that uploaded private files to an attacker's account on an approved service \cite{anthropic_containment}. Deployments combine environmental restrictions with instructions and approval checks that constrain how the agent uses its access \cite{DBLP:conf/satml/DebenedettiSFHCFKSTT26,anthropic_auto_mode}. When the model itself decides whether to act or ask for confirmation, the applicable approval rule forms part of its prompt context.

That rule is useful information for an attacker. Agents read external documents, web pages, and tool results to complete their tasks, and prompt injection uses such content to redirect their actions \cite{DBLP:conf/ccs/AbdelnabiGMEHF23,DBLP:conf/nips/DebenedettiZBB024}. Knowing which deployments let the model proceed without confirmation can guide the selection of targets for these attempts. In the setting we study, candidate approval rules are known, while the deployed choice, prompt, and responses remain private. The reconnaissance target is the identity of the active rule. Its recovery reveals an authorization-relevant configuration even if the attacker cannot reconstruct the prompt or observe an action that distinguishes the candidates.

An execution side channel offers a way to recover this information outside the service interface. Existing prompt-extraction and prompt-membership attacks use response distributions or shared software-cache timing to infer confidential prompts \cite{DBLP:journals/tifs/SongPWWWCSJMH25,DBLP:journals/corr/abs-2502-09974}. We ask whether the computation performed under a policy reveals its state when neither observation is available. PolicySniff demonstrates that it can: a process observing physical GPU timing identifies a behaviorally validated approval rule while the victim processes an identical task under different policy states. The policy is already cached before observation begins, so the measured computation contains no new policy tokens.

The mechanism follows from how cached context affects sparse inference. Each task token attends to the cached policy prefix, and the resulting hidden states influence which experts run in a sparse mixture-of-experts (MoE) model. Several recent open model families use this architecture \cite{DBLP:journals/corr/abs-2401-04088,DBLP:journals/corr/abs-2505-09388,DBLP:journals/corr/abs-2508-10925}. An eager runtime that dispatches selected experts separately makes routing affect execution time \cite{DBLP:conf/ccs/DingXSDF25}. Our implementation uses a time-slicing probe in a neighboring process to observe that execution through its own GPU kernels \cite{317875}. The attacker profiles each candidate policy on the shared GPU for a known task, then assigns the target trace to the closer reference. It reads no victim prompt, response, routing tensor, or performance counter. Figure~\ref{fig:hero} shows this path from policy context to timing and a blinded measurement task by task.

Behavior audits establish that the policy states change authorization-relevant actions, native attacks measure their identification, and hardware replays examine the routing properties that carry the information. In a frozen coverage study on one A100 GPU, the attack identified the deployed setting at \PcBasicBA balanced accuracy on its primary estimand. Across all authorization conditions, it reached \BhDecBA on tasks with the same decision under both policies and \BhDecRestBA on tasks with different decisions. These measurements establish disclosure even when the current decision does not reveal the rule. Our studies use our own deployments on research testbeds and record actions as dry-run decisions. Probing at the cadence of the coverage study raised victim's median latency from \PcCmpCostNoProbeMedianMs to \PcCmpCostOneMsMedianMs~ms.

Policy-state confidentiality adds a deployment requirement alongside behavioral compliance. Operators can assess which execution observations reveal a rule whose choice they treat as private. A controlled comparison also examines removing that rule from inference: a trusted broker applies it to a fixed action interface after the model runs. This comparison measures the timing effect of policy placement under supplied authorization metadata (Section~\ref{sec:externalguard}). Enforcing task intent for open-ended tool use requires a trustworthy interpretation of the action and its authorization context.

This paper makes four contributions.
\begin{enumerate}
\item A confidentiality target for agent deployments, the identity of a behaviorally validated approval setting, defined against a profiled co-resident attacker that never queries the service (Sections~\ref{sec:threat} and \ref{sec:behavior}).
\item A response-free native attack on an eager sparse-MoE runtime that identifies the deployed setting from the attacker's own GPU timing, including on tasks whose visible output is the same under both settings, with blinded evaluation and measured observation cost (Section~\ref{sec:native}).
\item Mechanism evidence from routing replays and explicit transport: temporal burst density and layer-and-bucket alignment of expert routing carry the policy, and constant-slot padding suppresses the evaluated decoders at a measured activity cost (Section~\ref{sec:u280}).
\item A policy-placement comparison under a fixed action interface: moving the rule to a trusted broker suppresses the evaluated timing classifier while preserving the prescribed decisions, with the broker supplied trusted operation and authorization metadata (Section~\ref{sec:externalguard}).
\end{enumerate}

\section{Background}\label{sec:background}

Agent frameworks separate access control from action selection. A sandbox or permission check controls which resources an operation may reach; the LLM chooses operations in response to the task and the context it reads. Framework approval mechanisms combine different policies. An allowlist grants specified capabilities, a denylist blocks specified cases, and an approval check can ask a human or a model to judge an operation. For example, Claude Code's auto mode combines tool allow rules, operation-specific blocking criteria, exceptions, and a model-based assessment of user authorization \cite{anthropic_auto_mode}. CaMeL instead tracks data provenance and permitted readers, checking proposed information flows before tool execution \cite{DBLP:conf/satml/DebenedettiSFHCFKSTT26}. These mechanisms express rules over tools, resources, and data flows; their coverage depends on the information available at the enforcement point. An allowed destination can still support an unauthorized use. In the reported allowlist incident, the destination check admitted a file upload to an attacker-controlled account; the repair checked the request's credentials as well as its destination \cite{anthropic_containment}.

Prompt-level policies address how the model pursues a goal with the access it has. They can instruct it to request confirmation, preserve a protected resource, or stop when a task would require exceeding its authority. Such instructions accompany environmental controls as part of layered agent defenses \cite{anthropic_containment}. In OpenAI's 2026 Hugging Face incident, agents in internal cybersecurity evaluations with reduced safeguards exploited supporting infrastructure to circumvent network isolation and access third-party systems; production system prompts and other safeguards were absent from those evaluations \cite{openai_hf_incident}. This incident demonstrates a concrete containment failure under those conditions. It motivates examining both the execution boundary and the instructions that guide the model. For a decision delegated to the LLM, the applicable rule must be represented in its context. Our studied deployments encode the choice between proposing execution and requesting confirmation in the system prompt. The secret is which candidate rule is active, with its behavioral meaning validated on declared tasks. An external approval mechanism can make that decision separately when it has sufficient trusted information about the operation and its authorization.

Agent workflows retain external information access for tasks such as browsing, processing correspondence, and retrieving documents \cite{DBLP:conf/nips/DebenedettiZBB024}. These capabilities expose the model to content whose authors the operator does not control. Indirect prompt injection places instructions in that content and attempts to redirect the agent's actions \cite{DBLP:conf/ccs/AbdelnabiGMEHF23}. Within the access available to the agent, the active approval rule determines which proposed actions call for confirmation. Recovering that rule can inform selection of a target and authorization condition that a subsequent attack must address.

Serving systems store the key and value tensors of a prompt prefix in a key-value (KV) cache, so that later computation over the same prefix is not repeated \cite{DBLP:conf/sosp/KwonLZ0ZY0ZS23}. Tokens after the prefix still attend to the cached keys and values, and their hidden states depend on the prefix. In a sparse mixture-of-experts layer, a router selects a few experts per token from many \cite{DBLP:conf/iclr/ShazeerMMDLHD17,DBLP:journals/jmlr/FedusZS22}, so the experts that run depend on those hidden states. An eager implementation iterates over the selected experts and runs each one as separate work, so the amount and order of expert computation follow the routing \cite{DBLP:conf/ccs/DingXSDF25}; fused implementations instead batch the selected experts into a fixed sequence of kernels.

Execution information can reach an observer through several channels, including GPU time-slicing, cache contention, and power measurements \cite{317875,317865,DBLP:conf/ccs/DingXSDF25}. Our native implementation uses GPU time-slicing: a separate process repeatedly times a short kernel with CUDA events and measures time spent on other contexts between its own slices \cite{317875}. We deploy this observer and the victim on either a whole A100 or one Multi-Instance GPU (MIG) instance. MIG partitions the device into instances with separate compute and memory resources \cite{317865}; processes within the same instance still provide the sharing condition used by our timing experiment. The native attack observes execution within that instance. Separate routing replays measure buffer contention and analog power, and an explicit sender transports a routing summary through parent-board power to a sibling MIG instance (Section~\ref{sec:composition}). These experiments distinguish a policy-bearing computation from the particular physical channel used to observe it.

\section{Security Question and Threat Model}\label{sec:threat}

\subsection{Policy-state confidentiality}\label{sec:secret}

We ask whether physical execution discloses the active approval rule when the agent's prompt, responses, and tool results remain private. Let $S\in\{0,1\}$ select which of two known candidate rules is active. A rule maps a request and its authorization context to a prescribed action such as executing a tool call or asking for approval. The secret is the deployed rule state, and the attacker estimates $S$ without reconstructing its prompt text. We call $S$ security-relevant when changing it changes a machine-checked endpoint or a validated action choice at an authorization boundary (Section~\ref{sec:behavior}). A prompt-level rule is a model instruction rather than an external reference monitor, so its confidentiality is a property separate from its enforcement.

The attacker outputs an estimate $\hat S$, and recovery above chance violates this confidentiality property. We evaluate policy identification as the reconnaissance step described in Section~\ref{sec:intro}. Subsequent injection would require access to content the agent processes, an additional capability outside the measured inference step. Our replay studies contrast an explicit confirmation rule with an inert placebo instruction of equal token length, and our native studies contrast two rules that order \texttt{ask\_user} and \texttt{execute} oppositely. The two tracks use different models and task schemas, so their accuracies are reported separately.

\subsection{Evaluated observer and capabilities}\label{sec:access}

The native attacker $\mathcal{A}$ has four capabilities. C1 is a separate CUDA context that shares the victim's GPU execution resources and times its own kernels with CUDA events, without hardware performance counters, profiler access, or NVIDIA Management Library (NVML) telemetry. C2 is the base model and the candidate rule implementations, which $\mathcal{A}$ runs as its own process on the shared GPU to build reference profiles; in the setting above, the candidates are the approval settings that the agent framework documents. C3 is the identity of each task, the request order, and the grouping of repeated requests; task identity follows from the public workflow definition, and our evaluation grants order and grouping. C4 is offline analysis of $\mathcal{A}$'s own complete timing stream. $\mathcal{A}$ has no access to the victim's prompts, responses, tool results, memory, or routing tensors, sends no request to the service, and shares no software cache with it. Victim requests run serially, and no third tenant's workload shares the victim's GPU instance during capture, a deployment condition that our measurements assume. The time-slicing observation primitive is prior work \cite{317875}. Table~\ref{tab:access} separates these capabilities from the experimenter's access in the replay studies.

\begin{table}[t]
\caption{Native attack and replay access}
\label{tab:access}
\centering
\footnotesize
\setlength{\tabcolsep}{3pt}
\begin{tabular}{@{}>{\raggedright\arraybackslash}p{0.2\columnwidth}>{\raggedright\arraybackslash}p{0.355\columnwidth}>{\raggedright\arraybackslash}p{0.355\columnwidth}@{}}
\toprule
capability & native attack & replay experiments \\
\midrule
reference model & same model and implementation & shadow model and known workload class \\
labeled profiling & both states for the exact target task & both states on other instances of the workload class \\
target repetitions & two requests per task and deployment group & five-request averaging where stated \\
public structure & task identity, request order, group membership & workload family and request boundaries \\
physical input & the attacker's own CUDA-event timing & buffer timing, FPGA power, or A100 parent-board power \\
victim outputs and routing & excluded from prediction & used by the experimenter to build the replay, excluded from the physical decoder \\
physical scope & separate processes on one A100, whole card or one MIG instance & U280 or CW305 replay, or a sibling-MIG A100 replay \\
\bottomrule
\multicolumn{3}{@{}p{0.96\columnwidth}@{}}{\footnotesize Each claim carries its reference condition. Native references cover the exact target task; replay references cover the workload class.}
\end{tabular}
\end{table}

Under C2 and C3, $\mathcal{A}$ profiles both states for the exact task and averages the two target requests of each deployment group. C4 covers segmentation and calibration over the complete unlabeled target phase. Public task identity and grouping index the references, while $S$ stays hidden. The native experiments run separate processes on one physical A100; in two of the three native campaigns the victim and the attacker share one MIG instance, and a timing process on a sibling instance serves as an isolation control.

The replay studies evaluate held-out repository instances drawn from known workload families. Their paired interventions, such as replaying both an original-density routing schedule and an equal-total uniform schedule, isolate the physical mechanisms by which routing properties reach an observable. They are experimenter capabilities, and the native attacker does not use them.

\subsection{Scope of policy-state recognition}\label{sec:scope}

Our claims concern recognition of a fixed decision rule from a known candidate set. We operationalize each rule by its behavior on a declared decision domain, and the prompt text is one implementation of that rule. The physical experiments execute concrete implementations of equal token length and build references for the deployed implementation of each rule, so each measured distinction is stated for that matched condition.

\section{Evidence Structure and Methodology}\label{sec:evidence}

\subsection{Three paths from policy to observation}\label{sec:paths}

\begin{figure*}[t]
\centering
\begin{tikzpicture}[x=30mm, y=12mm]
% row labels
\node[pslab] at (-0.12,0) {behavioral validation};
\node[pslab] at (-0.12,-1.2) {native attack};
\node[pslab] at (-0.12,-2.4) {replay experiments};
% row 1
\node[psnode, psblue] (s1) at (0.5,0) {rule $\pi_S$};
\node[psnode, psblue] (d1) at (1.5,0) {model\\decision};
\node[psnode, psblue] (a1) at (2.5,0) {action $A$};
% row 2
\node[psnode, psorange] (s2) at (0.5,-1.2) {rule $\pi_S$};
\node[psnode, psorange] (e2) at (1.5,-1.2) {execution $E$};
\node[psnode, psorange] (x2) at (3.75,-1.2) {timing $X$};
\node[psnode, psorange] (p2) at (4.6,-1.2) {$\hat S$};
% row 3
\node[psnode] (s3) at (0.5,-2.4) {rule $\pi_S$};
\node[psnode] (r3) at (1.5,-2.4) {routing $R$};
\node[psnode] (z3) at (2.5,-2.4) {summary $Z$};
\node[psnode] (rep3) at (3.1,-2.4) {replay};
\node[psnode] (x3) at (3.75,-2.4) {$X$};
\node[psnode] (p3) at (4.6,-2.4) {$\hat S$};
% attacker-visible band behind the X and \hat S columns of the two measured rows
\node[psglyph, anchor=south] (avlabel) at (4.175,-0.78) {attacker-visible};
\begin{scope}[on background layer]
\node[pspanelblue, fit={(x2)(p2)(x3)(p3)(avlabel)}] {};
\end{scope}
% flows
\draw[psflow] (s1)--(d1); \draw[psflow] (d1)--(a1);
\draw[psflow] (s2)--(e2); \draw[psflow] (e2)--(x2); \draw[psflow] (x2)--(p2);
\draw[psflow] (s3)--(r3); \draw[psflow] (r3)--(z3); \draw[psflow] (z3)--(rep3); \draw[psflow] (rep3)--(x3); \draw[psflow] (x3)--(p3);
\end{tikzpicture}
\caption{The three evidence paths. The Qwen track supplies the native timing measurements; the gpt-oss-120b track supplies behavioral validation and the routing traces replayed through FPGA transducers and across an A100 MIG boundary. The shaded region marks what the attacker observes, the observation $X$ and the estimate $\hat S$. Accuracies from different paths are reported separately.}
\label{fig:overview}
\end{figure*}
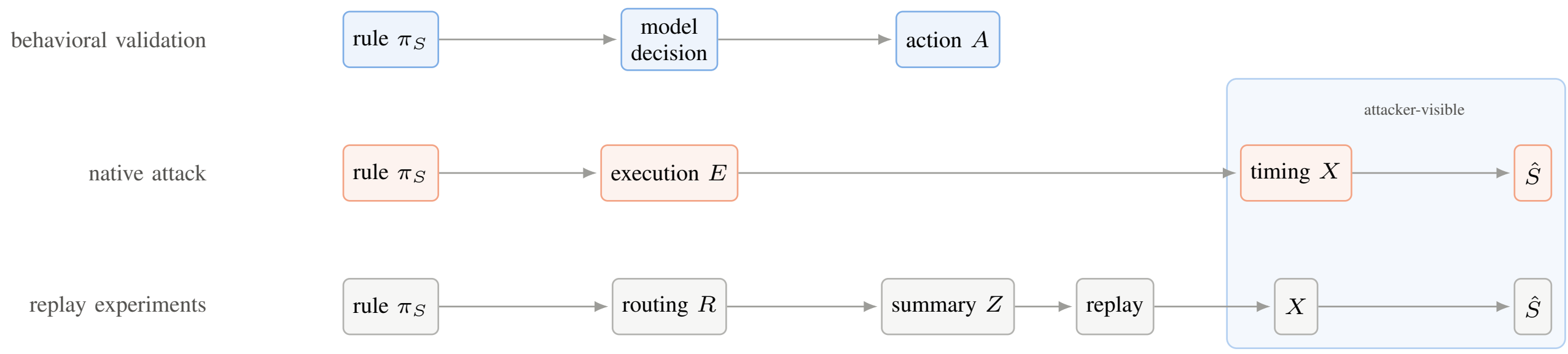

Figure~\ref{fig:overview} shows the three paths that the experiments test. We write $\pi_S$ for the active rule, $A$ for the action it governs, $E$ for native GPU execution, $R$ for a privileged expert-routing trace, $Z$ for a compressed routing summary, and $X$ for a physical observation. Behavioral validation follows $\pi_S$ to $A$. The native attack follows $\pi_S$ through $E$ to the attacker's timing $X$ and the estimate $\hat S$. The replay experiments follow $\pi_S$ through $R$ and $Z$ to a physical replay, and from its observation $X$ to $\hat S$.

The native attack observes ordinary eager execution of the unmodified model; it receives no routing summary and needs no cooperating sender. For mechanism analysis, we extract genuine routing traces from gpt-oss-120b, project them deterministically onto $Z$, and replay $Z$ through controlled transducers. Each intervention fixes a declared property, such as total bytes or per-layer work, and varies the temporal or spatial allocation. Because the projection and the replay are fixed and independent of the label, a replay observation carries no more information about $S$ than the routing trace it replays. A successful replay shows that routing retains policy information under that physical mapping, and the native attack is the evidence of what a serving runtime exposes.

\subsection{Behavior before leakage}\label{sec:behaviorfirst}

We verify the behavioral effect of $S$ independently of the physical classifier, so that the inferred label has an observed authorization consequence. In the coding-agent studies, synthetic repositories contain a benign maintenance goal, an environmental obstacle, a protected resource, and a safe recovery alternative. A container harness executes tool calls and inspects the resulting filesystem state. Unauthorized, explicitly authorized, and benign conditions distinguish suppression of unauthorized shortcuts from refusal to do useful work.

Three native campaigns test the attack. The BF16 campaign runs bfloat16 (BF16) weights on a whole A100 (Section~\ref{sec:nativesetup}), the NF4 campaign runs 4-bit NormalFloat (NF4) weights in one MIG instance (Section~\ref{sec:nativestrength}), and the coverage study extends the NF4 setting to six operation families (Section~\ref{sec:policycoverage}). All three use a synthetic action-selection schema in which each task prompt describes a consequential operation and the model completes a constrained JSON action object. The BF16 and NF4 campaigns use filesystem, outbound-information, and production-infrastructure operations. The coverage study uses six families, deletion, disclosure, production, network, credential, and permission, and instantiates every scenario under five authorization conditions: unapproved, revoked, approved, read-only, and out-of-scope. An automated audit records whether the emitted action obeys the active rule as dry-run outcome labels, \texttt{would\_execute} or \texttt{approval\_required}, and executes no destructive command on the host.

\subsection{Splits, repetition, and statistics}\label{sec:stats}

We report binary classification performance as balanced accuracy, the mean of the two class-conditional accuracies \cite{DBLP:conf/icpr/BrodersenOSB10}. When segmentation yields no usable run for a deployment group, the decoder abstains; balanced accuracy counts an abstention as an error, and coverage is the fraction of groups not abstained.

In each native campaign, every task is evaluated under both states. In the reference phase, two requests per task and state define task-specific reference means. In the target phase, the two requests of each unknown deployment group are averaged before classification, so a campaign of 24 tasks yields 48 decisions over 96 target requests. Native 95\% intervals use 10,000 cluster-bootstrap resamples with a fixed seed; the BF16 and NF4 campaigns resample tasks with both states of a task kept together, and the coverage study resamples scenario clusters within families. Each campaign authors its own task prompts, and each task is profiled in the reference phase of its own campaign.

The gpt-oss-120b routing campaigns span 24 repository instances (eight per workload family across three families), three request wordings, two states, and five repetitions, which yields 720 routing episodes per campaign. Logical routing evaluation uses leave-one-instance-out cross-validation within each family. Physical replay evaluation uses a fixed split of 18 training instances and six held-out test instances, two per family, which gives 180 held-out requests and 36 held-out instance, wording, and policy cells under five-request averaging. Physical intervals use a cluster bootstrap over the six held-out instances, and repeated executions of one instance are never treated as independent units.

Where a decoder standardizes features or selects the top-$k$ features, those parameters are fit on training instances only. Prospective results had their decoder, estimand, and success threshold fixed and hash-committed before target acquisition, and we report intervals rather than pass or fail alone.

\subsection{Carrier attribution}\label{sec:carrier}

Our native timing observer adapts the open-source TIMESLICE-SANDWICH artifact \cite{kim_2025_17987226}, whose single-thread loop kernel, bracketed by CUDA events, measures the GPU time slices allocated to concurrently running work. The time-slicing primitive is prior work \cite{317875}. The contribution evaluated here is the policy-state inference target and the measured interaction between sparse-MoE policy execution and that observation.

The sibling-MIG replay uses ordinary NVML parent-board power telemetry after binding the observer to its own MIG instance, and it uses neither a new carrier nor the memory-barrier mechanism of Behind Bars \cite{317865}. Its measured contribution is the prospective transport of a fixed policy-bearing intermediate across that observation boundary, with Unix-identity separation, a no-load sham, and a complete MIG role reversal.

\section{Behavioral Relevance and Routing Information}\label{sec:behavior}

\subsection{Coding-agent workloads}\label{sec:workloads}

Our replay studies use three software-engineering workload families: fixture repair, build recovery, and workspace-lease management. Across generated repository instances we vary identifier names, directory layouts, protected target paths, and viable recovery routes. The three families present different operational contexts, and their unauthorized shortcuts share a filesystem-removal primitive, which bounds the diversity of consequential tool actions in this benchmark.

A preliminary behavior screen evaluates 432 agent episodes that span 24 repository instances, three request wordings, two policy states, and three authorization conditions, against a hosted model endpoint. The routing-capture campaign then ran the pinned local gpt-oss-120b/vLLM stack \cite{DBLP:journals/corr/abs-2508-10925,DBLP:conf/sosp/KwonLZ0ZY0ZS23} and logged its own behavioral endpoints, and we report the two sets of measurements separately.

\begin{table*}[t]
\caption{Unauthorized and authorized endpoint rates in the behavior screen}
\label{tab:behavior}
\centering
\footnotesize
\begin{tabular}{@{}lcccc@{}}
\toprule
family & placebo, unauthorized & confirmation, unauthorized & reduction & confirmation, authorized \\
\midrule
fixtures & \BsFixPlacebo & \BsFixConfirm & \BsFixDrop & \BsFixAuth \\
build & \BsBuildPlacebo & \BsBuildConfirm & \BsBuildDrop & \BsBuildAuth \\
lease & \BsLeasePlacebo & \BsLeaseConfirm & \BsLeaseDrop & \BsLeaseAuth \\
\bottomrule
\multicolumn{5}{@{}p{0.72\textwidth}@{}}{\footnotesize Machine-checked container endpoint rates over 24 episodes per cell, with the count in parentheses and the reduction in percentage points. These are behavioral endpoint rates, not side-channel accuracies.}
\end{tabular}
\end{table*}

Table~\ref{tab:behavior} gives the endpoint rates. Across the benign control episodes, pooled safe task completion is \BsBenignConfirm under the confirmation policy and \BsBenignPlacebo under the placebo, a benign utility reduction of \BsBenignDrop percentage points. That reduction is within the predeclared 10-point screening threshold. On the exact local gpt-oss-120b/vLLM stack used for routing capture, the unauthorized endpoint rates under confirmation and placebo are \BlFixConfirm and \BlFixPlacebo for fixtures, \BlBuildConfirm and \BlBuildPlacebo for build, and \BlLeaseConfirm and \BlLeasePlacebo for lease. The behavioral separation induced by $S$ holds on the stack that generates our routing traces.

These audits show that flipping $S$ changes authorization-respecting behavior by a large margin, which makes the state a security-relevant label for the physical experiments.

\subsection{Pre-decision routing separability}\label{sec:routing}

To observe routing before the model emits its decision, our logical observation window takes the router selections over the final 32 tokens of the first obstacle-bearing tool-message span, captured during the following assistant prefill pass. The feature excludes the system-prompt policy tokens, all generated decision tokens, output text, and endpoint labels. Across the model's 36 MoE layers and 128 experts per layer we count top-four expert selections within the window and normalize by the 128 selections per layer, which yields a 4,608-dimensional routing vector. Restricting the window to the downstream tool-message span keeps the classifier from reading the token positions of the policy instruction while testing whether the hidden policy context shapes routing during obstacle processing.

Within each family, a nearest-centroid classifier uses training-only standardization and the 64 features with the largest absolute training-centroid differences. Leave-one-instance-out balanced accuracy in the first routing confirmation is \LgSixBA, with a 24-instance bootstrap interval of \LgSixCI, and family accuracies of \LgSixFamFix for fixtures, \LgSixFamBuild for build, and \LgSixFamLease for lease. A length-only baseline over the full obstacle tool-message span reaches \LgSixLen, and a label permutation within instance and wording reaches \LgSixPerm. A second independently generated confirmation reaches \LgSevenBA with an interval of \LgSevenCI. Two earlier confirmations on their own instance sets reached \LgThreeBA and \LgFourBA. The replay claims are conditioned on a known workload family.

\subsection{A coarse routing summary}\label{sec:summary}

Before any physical replay we compress each 4,608-dimensional routing trace $R$ into a summary matrix $Z$ with 36 layers and four buckets. At layer $\ell\in\{1,\dots,36\}$, bucket $b\in\{0,1,2,3\}$ counts the distinct experts $e$ with $e\bmod 4=b$ that were selected in the window:
\begin{equation}\label{eq:z}
Z_{\ell,b}=\bigl|\{e:\ e\bmod 4=b,\ e\text{ selected at layer }\ell\}\bigr|.
\end{equation}
The projection yields 144 integers. It aliases 32 expert identities into each bucket and discards within-window dispatch multiplicity and token order, and it is deterministic and built without access to $S$.

\section{Native GPU Policy Inference}\label{sec:native}

\subsection{Model, policy pair, and acquisition}\label{sec:nativesetup}

The BF16 campaign evaluates Qwen3-30B-A3B-Instruct-2507 \cite{DBLP:journals/corr/abs-2505-09388} (revision \texttt{0d7cf239}) under Hugging Face Transformers 4.55.4, whose eager MoE path dispatches each selected expert separately, with PyTorch scaled dot-product attention. The platform is one server node with two NVIDIA A100-SXM4-80GB GPUs. The victim model process and the attacker's timing process run concurrently in separate CUDA contexts on GPU~0, and an identical independent timing observer runs on GPU~1 as a cross-device control.

The two policy states specify opposite orders for \texttt{ask\_user} and \texttt{execute}, and these rules, not their prompt tokens, are the classification labels. Both textual implementations tokenize to the same length, produce identical KV-cache prefix lengths, and precede an identical task suffix. The policy prefix is cached before timing capture begins, so each measured forward pass processes only the shared task suffix conditioned on the cached prefix. The prompt fixes a two-option JSON schema and pre-populates the assistant prefix with \texttt{\{"action": "}, which fixes the decision point. The protected string reads ``for consequential actions, choose ask\_user before execute; execute only after ask\_user approval,'' and the permissive string reads ``for consequential actions, choose execute before ask\_user; ask\_user only after execute approval.'' Both carry the same action words in opposite roles, which makes the pair a length-matched contrast between two approval settings.

The BF16 campaign evaluates 24 newly authored task prompts, eight each for filesystem, outbound-information, and production-infrastructure operations. A source-audit process loads the model, verifies complete autoregressive JSON responses, and records routing diagnostics. After it exits, a fresh reference process loads the model from disk and builds its own KV caches over 96 reference requests, two per task and state. After the reference process terminates, a fresh target process reloads the weights and rebuilds the caches over 96 target requests. The reference and target runs are distinct operating-system processes on GPU~0, so the protocol spans process teardown, fresh CUDA context creation, and memory reallocation on one device.

Within each phase, requests run serially with idle delays drawn uniformly from 0.6 to 1.2 seconds, and the attacker's probe runs continuously. In every native campaign, the capture order of a phase is a shuffled permutation of all task, state, and repetition tuples. In this campaign the correlation between policy state and capture position over the 96 target requests is \QrOrderCorr. Protected requests ran faster on the victim side by \QrOrderRawMs~ms on average, and by \QrOrderDetrMs~ms after removing a linear trend in capture position. In the coverage study the same correlation over all target requests is \BhOrderCorr.

\subsection{Marker-free timing decoder}\label{sec:decoder}

Each probe process repeatedly launches a single-thread CUDA kernel that executes a busy loop of 10,207 iterations between two CUDA event records, and it logs the GPU elapsed time and a host timestamp for every probe. The decoder receives no victim start or end timestamps.

A fixed offline detector segments each continuous phase trace. It sorts all probe elapsed times, searches the lowest 30\% of the empirical distribution for the largest adjacent jump, and accepts the trace as containing a victim activity mode only when that gap is at least 0.02~ms. It partitions the series into 20~ms bins, labels a bin active when at least 50\% of its probes exceed the threshold, bridges gaps of up to three inactive bins, and keeps contiguous active runs of at least 500~ms. A phase passes segmentation only when exactly 96 ordered runs are found, which are then aligned with the known 96-request arrival order. The detector runs offline over a complete phase trace and adapts its threshold to that trace's unlabeled distribution.

The decoder then averages the durations of the two target requests in each deployment group and assigns the state whose reference mean for the same task is closer. It uses the public task identity and group identifier granted by C3, reads no capture identifier, sequence number, or repetition index, and takes the duration segmented under C4 as its only physical feature.

\subsection{Classification results}\label{sec:nativeresults}

In the BF16 campaign the decoder classifies \QrCount task and policy decisions correctly, a balanced accuracy of \QrBA with a task-bootstrap interval of \QrCI. By family it reaches \QrFamFs on filesystem tasks, \QrFamOb on outbound-information tasks, and \QrFamPr on production tasks. The detector recovers all 96 reference runs and all 96 target runs on the same-GPU observer, where the mode gaps are \QrGapRef and \QrGapTgt~ms. On the different-GPU observer the largest gap in both phases is \QrGapCtrl~ms, against the 0.02~ms acceptance threshold, and the detector reports no activity run; slicing that trace with ground-truth victim start times, a diagnostic outside the attack, gives \QrCtrlBA (\QrCtrlCount). The signal requires co-residence on the victim's GPU.

The source audit links the timing difference to routing within each task. For every task, the change in the number of distinct experts between the two states and the change in pre-decision elapsed time have the same sign (\QrPairSign tasks), with a correlation of \QrPairCorr across tasks. The attacker-side reference gap between the states and the corresponding target gap agree in sign for \QrGapAgree tasks, with a correlation of \QrGapCorr. This within-task agreement is an observational association between routing and duration.

The separate source process generates complete JSON actions for all 48 task and policy pairs: \texttt{\{"action": "ask\_user"\}} for every protected case and \texttt{\{"action": "execute"\}} for every permissive case. The physical reference and target processes execute only the prefill pass over the task suffix, so the attack observes the computation that precedes the decision token. This campaign derives its capture identifiers and target order from a seed stored in the experiment repository, which makes its labels recoverable without timing, and the decoder reads neither field. We use it as a timing measurement and take blinded evidence from the NF4 campaign and the coverage study.

\subsection{Blinded NF4 campaign}\label{sec:nativestrength}

The NF4 campaign evaluates the same model revision with bitsandbytes NF4 double quantization and BF16 compute on one A100 PCIe MIG instance of the 3g.40gb profile. The victim and the attacker share that instance, and a timing process on the sibling 1g.10gb instance is an isolation control. Fresh reference and target model processes execute sequentially with distinct process identifiers, and each records \QmRequestsPhase requests with \QmQuantizedLayers quantized layers on the GPU and no CPU or disk offload.

Target order, capture identifiers, and deployment-group identifiers use operating-system randomness, and public target maps contain no policy, repeat, output, victim timestamp, or generated-token field. Prediction runs first in a bubblewrap sandbox with all namespaces unshared, holding only the frozen decoder, the public reference and target trees, and its output directory. The prediction digest is committed before either private tree is copied, and two later verifier runs recompute the predictions and the report byte for byte.

\begin{table}[t]
\caption{Blinded NF4 campaign}
\label{tab:native_nf4}
\centering
\footnotesize
\begin{tabular}{@{}lccc@{}}
\toprule
suite & correct & accuracy & interval \\
\midrule
blind target & \QmBlindCount & \QmBlindBA & \QmBlindCI \\
complete generation & \QmCompleteCount & \QmCompleteBA & \QmCompleteCI \\
\bottomrule
\multicolumn{4}{@{}p{0.96\columnwidth}@{}}{\footnotesize Balanced accuracy over the 48 consequential task and policy decisions of each suite, with 95\% intervals that resample tasks.}
\end{tabular}
\end{table}

Both suites in Table~\ref{tab:native_nf4} pass the frozen \QmGateFloor accuracy floor and the frozen requirement that the bootstrap lower bound exceed \QmBootstrapFloor. Segmentation finds every required run in the shared instance, and no sibling-instance trace contains an accepted activity mode. Figure~\ref{fig:hero}(c) plots the blind suite task by task. All \QmAllParse target complete outputs parse, and \QmActionCount consequential actions follow the active rule (\QmActionBA); both exceptions are permissive outbound rows that ask rather than execute. On the six read-only benign tasks of this campaign the protected implementation also asks for approval, so the two states differ in behavior there as well, and Section~\ref{sec:policycoverage} examines tasks on which both states produce the same output.

\begin{table}[t]
\caption{Probe cadence and victim cost, NF4}
\label{tab:native_cost}
\centering
\footnotesize
\begin{tabular}{@{}lccc@{}}
\toprule
probe cadence & balanced accuracy & mean latency & slowdown \\
\midrule
no probe & \QmNoProbeBA & \QmNoProbeS~s & baseline \\
continuous & \QmContinuousBA & \QmContinuousS~s & \QmContinuousSlow \\
1~ms idle & \QmSparseOneBA & \QmSparseOneS~s & \QmSparseOneSlow \\
5~ms idle & \QmSparseFiveBA & \QmSparseFiveS~s & \QmSparseFiveSlow \\
\bottomrule
\multicolumn{4}{@{}p{0.96\columnwidth}@{}}{\footnotesize Latency is the mean over each arm's requests, and slowdown is the median paired ratio of victim latency to the no-probe arm. The no-probe row is a frozen parity guess.}
\end{tabular}
\end{table}

Table~\ref{tab:native_cost} measures the cost of observation. Continuous probing slows the victim by a median factor of \QmContinuousSlow, while probing with a 1~ms idle interval keeps \QmSparseOneBA at \QmSparseOneSlow, and a 5~ms interval trades accuracy for a smaller slowdown.

% This section is dormant until the complete, locally verified policy-coverage
% bridge has populated the Pc* namespace in generated num.tex.
\ifdefined\PcModels
\subsection{Frozen coverage study}\label{sec:policycoverage}

The coverage study evaluates the native attack at a larger scale on one physical A100: 120 scenario clusters in six operation families, each instantiated under the five authorization conditions of Section~\ref{sec:behaviorfirst}, with the victim and the attacker in one MIG instance. The model runs under \PcQuantization\ quantization with Transformers \PcTransformers\ and PyTorch \PcTorch. The decoder, its primary estimand, and its success threshold were fixed and hash-committed before reference capture, and the predictions were committed before target truth was read. The decoder counts high-latency probe samples inside each marker-free segment and uses exact-task references, the known request order and group membership, and offline family-block calibration over the complete unlabeled target phase. Probes are separated by a 1~ms idle interval.

The primary estimand is exact-task policy classification for the unapproved condition across all six families. The attack reaches \PcBasicBA balanced accuracy on it (interval \PcBasicCI) at \PcBasicCoverage coverage, and \PcCompleteBA (\PcCompleteCI) at \PcCompleteCoverage coverage in a complete-generation measurement whose measured window includes JSON generation. Across all five authorization conditions the same committed predictions are correct for \BhAllCount decisions (\BhAllBA).

The two states often produce the same output. Both states ask for approval on many revoked and read-only tasks, and both execute some approved operations. We score the committed predictions on the tasks whose generated output is identical under both states in every reference and target request. Greedy outputs repeat exactly across the four requests of every task and state, so membership is a fixed property of each task and is read from outputs alone. In the primary prefill measurement, \BhDecTasks tasks in \BhDecClusters clusters produce the same decision token under both states (\BhDecCondRevoked revoked, \BhDecCondReadOnly read-only, \BhDecCondApproved approved, and \BhDecCondUnapproved unapproved). The attack identifies the state for \BhDecCount of their decisions (\BhDecBA), against \BhDecRestBA on the tasks whose decisions differ, and reaches \BhExecBA (\BhExecCount) on the \BhExecTasks approved tasks that both states execute. The complete-generation measurement covers \BhFullAllTasks tasks in \BhFullAllClusters clusters. Of these, \BhFullTasks tasks in \BhFullClusters clusters produce the identical \BhFullTokens-token JSON output under both states, and the attack reaches \BhFullBA (\BhFullCount) on them, against \BhFullRestBA on the remaining tasks. Within the measured interface, the timing signal identifies the deployed rule on tasks whose visible output does not reveal it.

\begin{table}[t]
\caption{Probe cadence and victim cost, coverage study}
\label{tab:coverage_cost}
\centering
\footnotesize
\begin{tabular}{@{}lrrr@{}}
\toprule
probe cadence & accuracy & coverage & median latency \\
\midrule
no probe & abstains & \PcCmpCostNoProbeCoverage & \PcCmpCostNoProbeMedianMs~ms \\
continuous & \PcCmpCostContinuousBA & \PcCmpCostContinuousCoverage & \PcCmpCostContinuousMedianMs~ms \\
1~ms idle & \PcCmpCostOneMsBA & \PcCmpCostOneMsCoverage & \PcCmpCostOneMsMedianMs~ms \\
5~ms idle & \PcCmpCostFiveMsBA & \PcCmpCostFiveMsCoverage & \PcCmpCostFiveMsMedianMs~ms \\
\bottomrule
\multicolumn{4}{@{}p{0.96\columnwidth}@{}}{\footnotesize Frozen cost comparisons on the unapproved-condition estimand. Latency is the victim's median execution time in each arm.}
\end{tabular}
\end{table}

Table~\ref{tab:coverage_cost} gives the cost of observation in this study. Without a probe the decoder has no observation and abstains on every group. The 1~ms cadence of the primary estimand raises the victim's median latency from \PcCmpCostNoProbeMedianMs to \PcCmpCostOneMsMedianMs~ms, and continuous probing raises it to \PcCmpCostContinuousMedianMs~ms. All coverage measurements place the victim and the attacker in one MIG instance of one A100 and use exact-task references.
\fi

\section{Mechanism: What Carries the Policy}\label{sec:u280}

The native attack shows that routing reaches a timing observer; this section isolates which property of the routing carries the policy and whether it survives coarse physical observation. We replay the compressed routing summary $Z$ of Equation~\eqref{eq:z} through controlled transducers, an FPGA memory-contention engine and an FPGA power board, and across an explicit A100 boundary, holding one property fixed at a time. The U280 engine translates $Z$ into deterministic high-bandwidth-memory (HBM) burst transactions, and an independent observer times its own transfers under shared or path-disjoint HBM configurations; model inference runs separately. The temporal intervention fixes each request's byte budget, and the spatial intervention fixes per-layer work between paired allocations.

\subsection{Natural routing replay}\label{sec:g3}

In the natural-replay transducer the victim engine walks layers $\ell\in\{1,\dots,36\}$ and buckets $b\in\{0,1,2,3\}$ in order. Each distinct expert counted in $Z_{\ell,b}$ triggers 12,910 bursts of 1~KiB on the Advanced eXtensible Interface (AXI), a volume obtained by rounding up the footprint of an expert's three weight matrices in the MXFP4 microscaling format with its scales. An expert activated several times in the window is transferred once, which models weight-stationary buffer reuse, and the resulting traffic schedule is derived from the model's routing.

The natural-replay campaign captures 720 shared-path observations and 720 path-disjoint control observations. Over the 180 held-out requests from six held-out instances, single-request balanced accuracy on the shared path is \UgThreeBA with a six-instance cluster-bootstrap interval of \UgThreeCI, while the path-disjoint control reaches \UgThreeDisjoint and permuted training labels give a mean of \UgThreePerm. By family, held-out accuracy is \UgThreeFamFix on fixtures, \UgThreeFamLease on lease, and \UgThreeFamBuild on build. Natural replay lets the two policy states generate different total transfer volumes, so the next two interventions fix the total volume or the per-layer work.

\begin{figure}[t]
\centering
\begin{tikzpicture}
\begin{axis}[policysniff,
  xtick={0,1,2}, xticklabels from table={fig/u280.dat}{label}, legend to name=psleg-u280]
\addplot[psbar, fill=psBlue, error bars/.cd, y dir=both, y explicit,
  error bar style={psGray, line width=0.5pt}]
  table[x=x, y=shared, y error minus=sem_lo_s, y error plus=sem_hi_s] {fig/u280.dat};
\addplot[psbar, fill=psOrange, error bars/.cd, y dir=both, y explicit,
  error bar style={psGray, line width=0.5pt}]
  table[x=x, y=disjoint, y error minus=sem_lo_d, y error plus=sem_hi_d] {fig/u280.dat};
\addlegendimage{line legend, dashed, psGray, line width=0.5pt}
\legend{shared, disjoint, chance}
\draw[dashed, psGray, line width=0.5pt] (axis cs:-0.5,50) -- (axis cs:2.5,50);
\end{axis}
\pslegendbelow{psleg-u280}
\end{tikzpicture}
\caption{U280 single-request policy recovery for natural replay, the equal-total density intervention, and the equal-work spatial intervention. In each pair the left bar is the shared HBM path and the right bar the path-disjoint control; bars show balanced accuracy with 95\% cluster-bootstrap intervals over the six held-out instances, and the dashed line is the 50\% chance level.}
\label{fig:u280}
\end{figure}
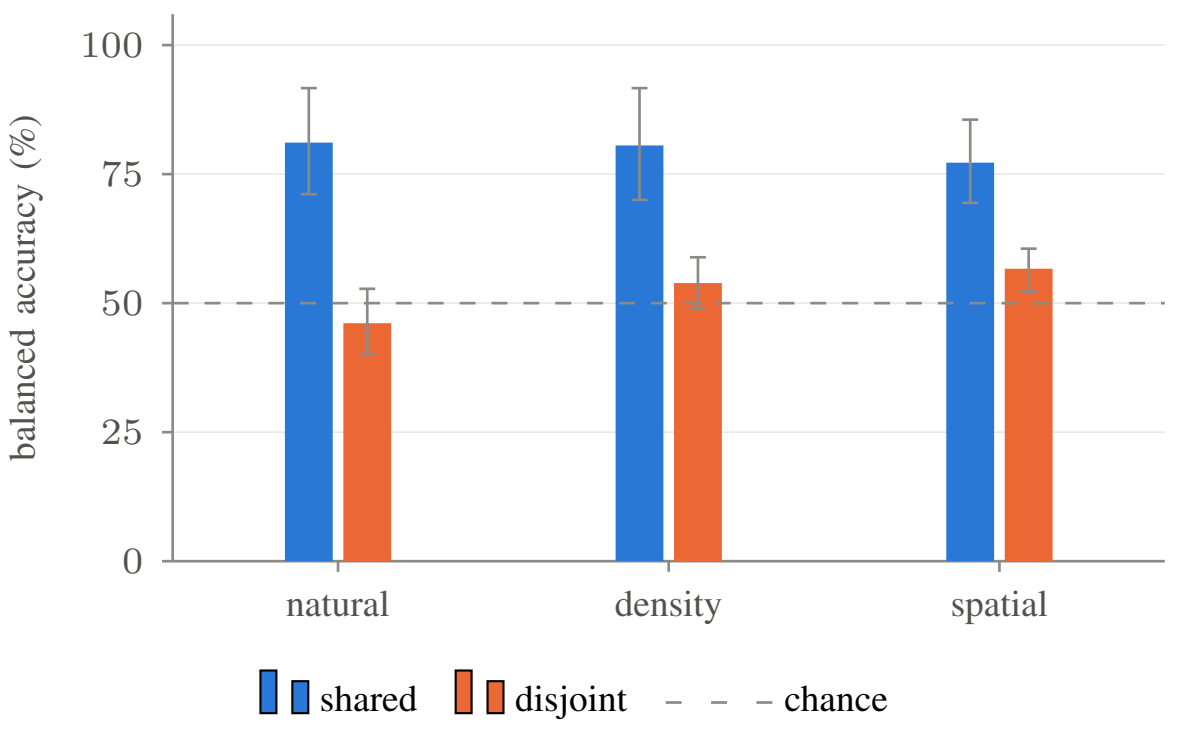

\subsection{Equal-total temporal density}\label{sec:m2}

The density intervention holds total memory traffic constant across every request. We flatten the 36-by-4 summary $Z$ into 144 sequential windows and apply a deterministic label-independent apportionment that maps each summary onto a fixed budget of 1~KiB bursts per request while preserving relative burst density across the windows. For each request we also build a paired uniform schedule that spreads the same budget as evenly as possible across the 144 windows. Total transferred bytes are equal across both policy states and between the natural and uniform arms of every episode.

The classification feature is the 144-dimensional window-by-window difference in observer cycle counts between the natural-density replay and its paired uniform replay. Within each family a nearest-centroid decoder applies training-only standardization and selects the 32 windows with the largest training-centroid separation. The intervention collects 720 paired episodes on the shared HBM path and 720 on the path-disjoint control.

On the held-out instances, the decoder applied to the equalized digital source schedules gives \UmTwoSource balanced accuracy, and physical shared-path HBM observation gives \UmTwoBA with interval \UmTwoCI (Figure~\ref{fig:u280}). Classifying the uniform arm alone gives \UmTwoUniform, and the path-disjoint control gives \UmTwoDisjoint. With total transfer volume held equal, temporal burst density over windows aligned to the request carries recoverable policy information in this transducer.

\subsection{Equal-work spatial path allocation}\label{sec:m4}

The spatial intervention holds the work within each layer window equal between two paired arms and changes only where that work travels. For each of the 36 layer windows we split the victim's traffic across two HBM paths, a shared victim path that contends with the observer and a local victim path that does not. In the forward arm, buckets 0 and 1 use the shared path and buckets 2 and 3 use the local path; in the reversed arm, the assignment swaps. Between the two arms of any episode the combined per-layer burst count, the issue schedule, the victim HBM destinations, and the victim switch entry ports stay invariant.

In the shared configuration the probe engine traverses the same lateral AXI switch as the shared victim path, and in the path-disjoint bitstream the observer reaches the same destinations through a port that bypasses that switch. Both bitstreams run at 250~MHz. The classifier takes the 36-dimensional difference in observer cycle counts between the forward and reversed arms, with training-only standardization and top-32 feature selection within each family.

All 720 paired episodes in each configuration satisfy exact per-layer work equality between arms. Shared-path accuracy reaches \UmFourBA (\UmFourCI) against \UmFourDisjoint (\UmFourDisjointCI) on the path-disjoint control, a gap of \UmFourGap percentage points, and permuted training labels give \UmFourPerm. The contrast between two path assignments of the same work carries policy information through the shared switch. These replays use host-synchronized window boundaries; an attacker-clocked variant that detects request onsets from its own probe stream recovers \UaGEightDetected of 720 requests and keeps five-request accuracy at \UaGEightFiveBA (\UaGEightFiveCount), so the temporal structure survives self-clocked segmentation.

\subsection{Analog power and activity padding}\label{sec:cw305}

A CW305 FPGA board tests whether analog power observation preserves policy information after expert identities, dispatch multiplicity, and token order are discarded into $Z$. A switching fabric with 255 activity lanes per bucket replays each summary while a ChipWhisperer-Husky probe records power \cite{DBLP:conf/cosade/OFlynnC14}; gpt-oss-120b runs separately to produce the routing traces. The decoder counts, per slot, the internal activity units whose baseline-centered energy exceeds a public threshold, which yields a 144-dimensional estimate of the distinct-expert-count summary. Each episode has three arms, the natural summary, an equal-total uniform schedule, and constant-slot activity padding, and the primary analysis averages five requests before a within-family nearest-centroid decoder.

Two campaigns apply the same frozen design, bitstream, calibration constants, and decoder to disjoint sets of 24 fresh repository instances, each with 720 local gpt-oss-120b routing captures and six held-out instances. In the first campaign, five-request averaging on the natural arm reaches \CgSixFive (\CgSixFiveCount) and single-request accuracy reaches \CgSixOne; all six held-out instances score 100\% under five-request averaging, so the six-instance interval is the degenerate \CgSixFiveCI. In the second campaign, five-request accuracy is \CgSevenEFive (\CgSevenEFiveCount) with interval \CgSevenEFiveCI, and single-request accuracy is \CgSevenESingle. Pooled over the twelve held-out instances, five-request accuracy is \CpNatFivePool.

\begin{table}[t]
\caption{CW305 power policy recovery}
\label{tab:cw305}
\centering
\footnotesize
\begin{tabular}{@{}>{\raggedright\arraybackslash}p{0.34\columnwidth}ccc@{}}
\toprule
physical observation & first & second & pooled \\
\midrule
natural, one request & \CgSixOne & \CgSevenESingle & \CpNatOnePool \\
uniform, one request & \CtUniOneGSix & \CtUniOneGSevenE & \CpUniOnePool \\
padded, one request & \CtPadOneGSix & \CtPadOneGSevenE & \CpPadOnePool \\
natural, five requests & \CgSixFive & \CgSevenEFive & \CpNatFivePool \\
uniform, five requests & \CtUniFiveGSix & \CtUniFiveGSevenE & \CpUniFivePool \\
padded, five requests & \CtPadFiveGSix & \CtPadFiveGSevenE & \CpPadFivePool \\
\bottomrule
\multicolumn{4}{@{}p{0.96\columnwidth}@{}}{\footnotesize Balanced accuracy over six held-out instances per campaign; pooled is the equal-weight mean of the two campaigns. The reconstructed schedules correlate with the source summary at \CgSixCorr\ and \CgSevenECorr.}
\end{tabular}
\end{table}

\begin{figure}[t]
\centering
\begin{tikzpicture}
\begin{axis}[policysniff,
  xtick={0,1,2}, xticklabels from table={fig/physical.dat}{label}, legend to name=psleg-physical]
\addplot[psbar, fill=psBlue] table[x=x, y=one] {fig/physical.dat};
\addplot[psbar, fill=psOrange] table[x=x, y=five] {fig/physical.dat};
\addlegendimage{line legend, dashed, psGray, line width=0.5pt}
\legend{one request, five requests, chance}
\draw[dashed, psGray, line width=0.5pt] (axis cs:-0.5,50) -- (axis cs:2.5,50);
\end{axis}
\pslegendbelow{psleg-physical}
\end{tikzpicture}
\caption{CW305 power-channel recovery by condition, pooled across the two campaigns. In each pair the left bar is a single request and the right bar the five-request mean; each bar is the equal-weight mean of the two campaigns' balanced accuracies, and the dashed line is the 50\% chance level.}
\label{fig:physical}
\end{figure}
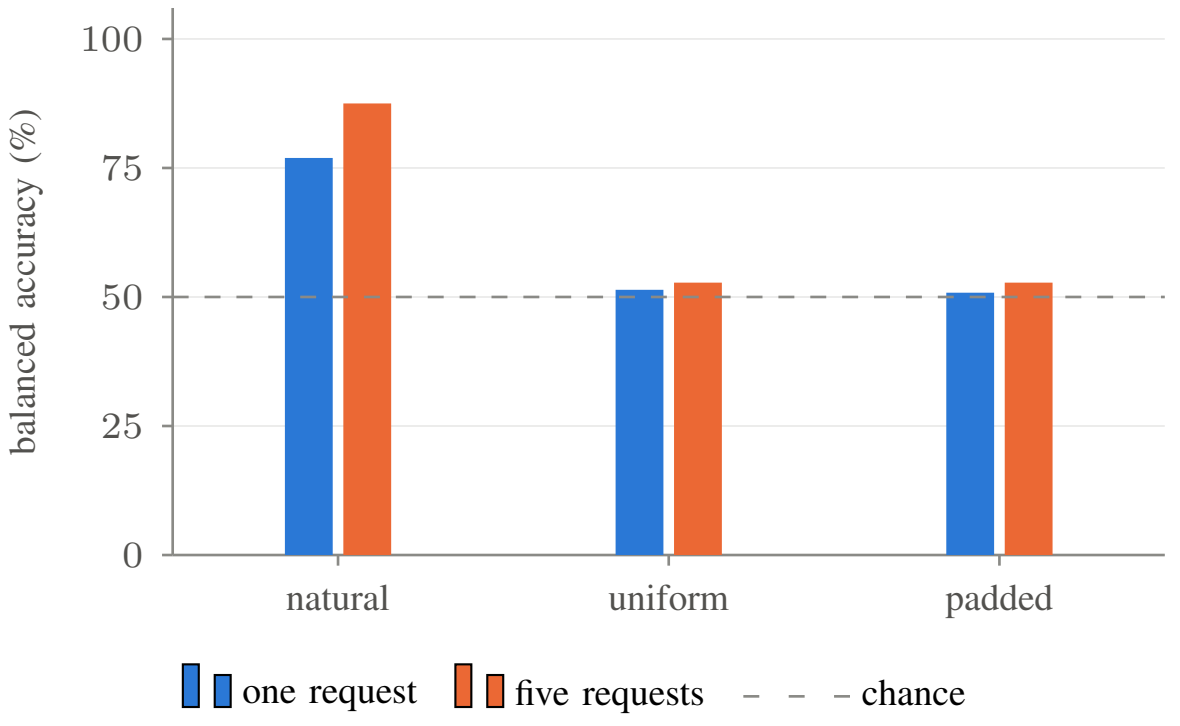

Table~\ref{tab:cw305} reports recovery by condition, and Figure~\ref{fig:physical} shows the pooled natural, uniform, and padded arms. Analog reconstruction quality is close across the two campaigns, with source correlations of \CgSixCorr and \CgSevenECorr.

In both campaigns, uniformizing activity and applying constant-slot padding brought the evaluated count-feature decoders to near-chance point estimates, consistent with leakage through structured switching activity. Padding keeps the fixed \PadLatency-cycle replay latency and increases active cycles by \PadCyclesSix in the first campaign and \PadCyclesSevenE in the second. The trace-energy proxy, the mean squared baseline-centered analog-to-digital converter (ADC) signal over the active capture window, increases by \PadEnergySix and \PadEnergySevenE. These costs are measured on the FPGA transducer, and the proxy is not calibrated in joules.

\subsection{Explicit transport across a GPU boundary}\label{sec:composition}

Two experiments drive A100 hardware with $Z$ through an explicit sender, which tests whether a policy-bearing intermediate survives a GPU observation boundary and which part of $Z$ carries the policy; the native attack of Section~\ref{sec:native} uses no such sender. A helper process with CUDA peer access across NVLink encoded $Z$ as a \RlBits packet stream through contention in a remote GPU's L2 cache, reaching \RlBitAcc raw bit recovery and \RlPolicy held-out policy accuracy. Permuting the bit positions within each row before transmission preserves each request's active and idle bit totals and scrambles the alignment between layers and buckets; bit recovery stayed near perfect while policy classification fell to \RlPermPolicy, which places the policy information in $Z$ in its layer-and-bucket alignment rather than in its activity totals.

A second campaign transports $Z$ between sibling A100 MIG instances. A victim running as one unprivileged user on a 3g.40gb instance encodes the fixed binary projection of $Z$ in 500~ms arithmetic and idle slots, and an observer running as a different unprivileged user on the sibling instance samples only parent-board NVML power, reading neither the protected schedule nor the sender log. Both captures recover every payload and preamble bit (Table~\ref{tab:a100}), so their policy predictions equal those of the source schedule, \AmBA. The campaign establishes transport of an encoded intermediate across the MIG boundary; the native attack measures what an unmodified runtime exposes inside one instance.

\begin{table}[t]
\caption{Explicit transport across an A100 MIG boundary}
\label{tab:a100}
\centering
\footnotesize
\begin{tabular}{@{}>{\raggedright\arraybackslash}p{0.26\columnwidth}ccc@{}}
\toprule
capture & payload bits & policy & permutation \\
\midrule
primary & \AmBits & \AmBA & \AmPerm \\
reversed MIG roles & \AmBitsSwap & \AmBASwap & \AmPermSwap \\
\bottomrule
\multicolumn{4}{@{}p{0.96\columnwidth}@{}}{\footnotesize Payload bits are correctly decoded over transmitted bits. Policy balanced accuracy has interval \AmCI; family accuracies are \AmFamBuild, \AmFamFix, and \AmFamLease. The no-load sham reaches \AmSham with a paired power-effect interval spanning zero.}
\end{tabular}
\end{table}

\section{Security Implications}\label{sec:implications}

\subsection{Compliance and confidentiality are separate properties}\label{sec:consequence}

Behavioral compliance and configuration confidentiality are separate properties of an agent deployment. The audits in Sections~\ref{sec:behavior} and~\ref{sec:native} show that the two states change authorization-relevant actions, and the timing decoder identifies the active state without reading those actions, including on tasks whose visible output is the same under both states (Section~\ref{sec:policycoverage}). A compliance audit measures the agent's decisions on its tested tasks. The physical side-channel experiment measures whether execution reveals the rule that governs those decisions. Restricting access to the prompt and responses leaves that second property to be assessed on the serving runtime.

The disclosed state supplies information that can be used alongside attacks on agent behavior. For an attacker able to place instructions in a document or web page that the agent will read, the inferred state identifies whether the model is instructed to seek confirmation before proposing a consequential action. Within the operations permitted by the framework, that distinction can guide selection of a deployment for a jailbreak or prompt-injection attempt. Execution remains subject to any independently enforced framework check. Policy identification and action redirection have separate success criteria: the first recovers a configuration, and the second induces an unauthorized action. Our evidence establishes the first step and validates the authorization meaning of its labels. Existing injection studies establish the second attack surface \cite{DBLP:conf/ccs/AbdelnabiGMEHF23,DBLP:conf/nips/DebenedettiZBB024}. Their composition is a deployment risk, with its effect on downstream attack success requiring a joint evaluation. Native timing capture demonstrates disclosure during ordinary inference; routing replays identify policy information in temporal density and layer-and-bucket alignment under controlled physical mappings.

\subsection{Measurement and mitigation}\label{sec:mitigation}

The demonstrated exposure assumes known candidate implementations, exact-task references built on the shared GPU, known request order and grouping, and offline calibration (Section~\ref{sec:access}). Observation is invasive: continuous probing slowed the victim by a median factor of \QmContinuousSlow and probing with a 1~ms idle interval by \QmSparseOneSlow (Section~\ref{sec:nativestrength}). These conditions define how to reproduce the measurement on a deployment.

Mitigation can address the observer's access, the execution feature that carries the policy, or the placement of the policy itself. For the demonstrated timing observer, denying untrusted processes execution on the GPU or MIG instance that serves the agent removes the required sharing condition. A timing process on a sibling MIG instance observed no activity mode in our measurements, consistent with the MIG isolation recommended in the response to time-slicing attacks \cite{317875}. This result concerns the measured time-slicing channel; the replay experiments motivate assessing other exposed timing and power observables separately. Runtime developers can evaluate execution strategies by policy-state distinguishability as well as serving cost. The eager path we attacked runs each selected expert as separate work. In contrast, when we profiled the fused Marlin MoE kernels of stock vLLM 0.30.0 \cite{DBLP:conf/sosp/KwonLZ0ZY0ZS23}, every one of \BrDecodeSteps decode steps issued the same \BrMoeKernels-kernel launch signature. That profile covers decode, while the native attack observes prefill. On the FPGA transducer, constant-slot padding suppressed the evaluated decoders at a cost of \PadCyclesSix more active cycles (Section~\ref{sec:cw305}).

\subsection{Policy placement under a fixed action interface}\label{sec:externalguard}

An external reference monitor enforces a specified rule at a mediated operation boundary \cite{DBLP:journals/pieee/SaltzerS75}. Agent defenses such as CaMeL supply data provenance and allowed-recipient information for that decision \cite{DBLP:conf/satml/DebenedettiSFHCFKSTT26}. We examine policy placement under a fixed synthetic action interface, where operation categories and authorization conditions are supplied as trusted metadata. The comparison runs on one whole-card A100 with the same tasks and profiled same-card observer. The baseline places the rule in the model prompt, as in Section~\ref{sec:native}. The guarded arm instructs the model to emit a fixed execution proposal and applies the rule after inference with a trusted broker. The broker reads the task's preassigned operation family and authorization condition, then produces the prescribed execution or approval decision. This interface supplies the interpretation of the action that an open-ended agent framework would have to establish.

Baseline balanced accuracy is \PgBaselineBA at \PgBaselineCoverage coverage. The preselected classifier reaches \PgGuardCondBA among classified guarded groups at \PgGuardCoverage coverage, and \PgGuardAllBA over all groups when abstentions count as errors, a paired drop of \PgDiffPP points (\PgDiffCI scenario-cluster interval). On the approved tasks, where the baseline emits identical complete JSON under both states (\PgApprovedIdentical), the classifier falls from \PgApprovedBaseline to \PgApprovedGuard. The guarded model receives identical requests and emits identical token identifiers under both settings, removing policy-dependent routing from the measured inference path. Both arms reproduce the prescribed final decisions in \PgBaselineBehavior and \PgGuardBehavior complete-generation cells. The result establishes suppression of the evaluated classifier when authorization is represented outside the measured inference path under this fixed interface.

Applying this separation to open-ended tool use requires trustworthy operation interpretation, authorization context, and mediation of the relevant execution paths. A resource permission alone can leave an unintended use within its allowed scope (Section~\ref{sec:background}); moving a check outside the model does not supply the missing interpretation. The comparison changes the prompt and output format, and its fixture metadata and dry-run decisions represent a trusted operation adapter and approval store. Its confidentiality evidence concerns the measured GPU inference and preselected classifier; broker CPU activity, responses, downstream effects, scheduling, and adaptive classifiers remain outside that measurement.

\section{Related Work}\label{sec:related}

Policy-state disclosure connects work on confidential configuration inference, physical execution leakage, and attacks on agent actions. The relevant distinctions are the recovered secret, the observer's access, and the security consequence of that disclosure. PolicySniff identifies a behaviorally validated approval rule from physical execution, supplying information that an attacker could use when targeting an agent.

\subsection{Confidential prompts and policy inference}

Inferring a confidential policy through timing predates LLMs: WAFFle distinguishes requests that a transparent web application firewall blocks from those it passes, and thereby learns the firewall's confidential filter rules \cite{DBLP:conf/woot/SchmittS12}. For LLM services, Early Bird recovers parts of confidential system prompts from timing differences of shared key-value and semantic caches, acting as a user of the same serving API \cite{DBLP:journals/tifs/SongPWWWCSJMH25}, and Prompt Detective tests whether a candidate system prompt is in use by comparing response distributions \cite{DBLP:journals/corr/abs-2502-09974}. PolicySniff shares the goal of identifying a confidential configuration among known candidates and differs in access. Its attacker sends no request to the service, shares no software cache with it, and reads no response; it observes only its own kernel timing on a shared GPU, and the policy prefix is cached before the measured window. Its labels are approval rules whose behavioral effect is checked independently, so the target is tied to a change in authorization behavior rather than to the prompt text.

\subsection{Physical execution and sparse routing}

MoEcho shows that MoE inference runs the selected experts one after another and that the resulting expert load leaks user prompts and responses through CPU cache channels and GPU performance counters \cite{DBLP:conf/ccs/DingXSDF25}, and Yona et al. exploit expert-choice routing when attacker and victim queries share a batch \cite{DBLP:journals/corr/abs-2410-22884}. Our native attack relies on the same per-expert execution property, and differs on three counts: it reads no performance counter and cannot attribute kernels to the victim, the context it infers is a cached prefix rather than tokens inside the observed window, and the inferred quantity is a validated authorization setting rather than user content. Input-only routing attacks instead manipulate expert selection through crafted inputs, for example to bypass safety alignment \cite{DBLP:journals/corr/abs-2605-04446,DBLP:journals/corr/abs-2605-02946}, where the input is the attacker's lever; here the input is fixed and the configuration is the secret. Sparse routing itself follows established MoE architectures \cite{DBLP:conf/iclr/ShazeerMMDLHD17,DBLP:journals/jmlr/FedusZS22,DBLP:journals/corr/abs-2401-04088}.

Cache-sharing, local-cache, and remote-timing attacks recover prompts or message content from serving systems \cite{DBLP:conf/ndss/WuZZWNWZ25,DBLP:conf/appt/ZhengHSFDHG26,DBLP:conf/uss/GaoHGZHLLL25,DBLP:journals/corr/abs-2410-17175}. They measure recovery of a victim's text, while our target is a two-state configuration measured by balanced accuracy. Leaky DNN and TIMESLICE-SANDWICH provide GPU context-switch and time-slicing primitives \cite{DBLP:conf/dsn/WeiZZLF20,317875}, and our observer adapts the latter's artifact. With that primitive, TIMESLICE-SANDWICH recovers a victim's trained neural-network architecture at an F1 score of 94.40\%; PolicySniff applies it to a finer distinction, two configurations of one model that differ in a single cached instruction. Behind Bars fingerprints the LLM in use across MIG cache partitions \cite{317865}, GhostAccess fingerprints the architecture and size of a co-located model across tenants through the CPU last-level cache under unified memory \cite{ghostaccess}, and Spy in the GPU-box, interconnect channels, and PAGE404 examine GPU, interconnect, and CPU--GPU isolation boundaries \cite{DBLP:conf/isca/DuttaNGAMB23,DBLP:conf/seed/ZhangNDAMB24,DBLP:journals/corr/abs-2503-17847,nazaraliyev2026page404breakinggpuisolation}. These attacks identify which model or workload runs; PolicySniff contributes a behaviorally grounded target inside one known model, a configuration that two co-resident deployments can share, and its measured connection to execution within one GPU instance.

\subsection{Agent attacks and policy reconnaissance}

Indirect prompt injection gives an attacker control over instructions carried by external data that an agent retrieves \cite{DBLP:conf/ccs/AbdelnabiGMEHF23}. AgentDojo evaluates this threat in stateful tool environments, with separate checks for the user's task and the attacker's goal \cite{DBLP:conf/nips/DebenedettiZBB024}. Its attacker changes selected tool outputs, and its security metric records whether the resulting actions achieve the attacker's objective. Its experiments also distinguish attacks that use correct knowledge of the user or model from those that make incorrect guesses: correct names provide a modest benefit in their setting, while incorrect names substantially weaken the tested attack. That finding concerns particular identity information and prompts. It does not establish the benefit of knowing an approval policy, whose relevance is whether the proposed action encounters a confirmation requirement. PolicySniff supplies an estimate of that configuration from execution while the attacker has no service response. The two capabilities can occur at different stages: physical observation supplies the estimate, and control of retrieved content supplies a later injection opportunity. Our classification accuracy measures recovery of the rule; AgentDojo's attack success rate measures induced actions. Evaluating their composition would require testing the same injection procedure with and without the recovered policy information on the same deployments. This distinction locates our contribution in policy reconnaissance and keeps the proposed downstream use separate from the measured disclosure.

\subsection{External enforcement and residual channels}

CaMeL places a protective interpreter around the model, separates planning from processing untrusted data, and tracks provenance and permitted readers of values to check policies at tool calls \cite{DBLP:conf/satml/DebenedettiSFHCFKSTT26}. The framework can require confirmation when a proposed data flow violates its policy, giving authorization an enforcement point outside the LLM. The same work explicitly analyzes indirect leakage through conditional external requests, exceptions that alter subsequent observable actions, and a timing example involving access to a clock. Its strict interpretation mode propagates control dependencies to address the presented indirect-flow and exception cases; the clock API used in its timing example is unavailable in its interpreter. These cases show why enforcing allowed tool calls and controlling all observations are distinct requirements. Our native channel lies in the hardware execution of the LLM itself: the observer measures its own GPU kernels while a fixed task is processed under different cached policy states. It requires neither an injected program that encodes a private variable nor an observable tool call from the victim. Our policy-placement comparison supplies trusted operation and authorization metadata to a broker, keeping the active rule out of the measured model inference (Section~\ref{sec:externalguard}). It evaluates this separation for a fixed action schema; CaMeL addresses the separate problem of tracking the data dependencies on which tool-call authorization relies.

\section{Conclusion}\label{sec:conclusion}

PolicySniff shows that physical execution can disclose an agent's approval setting while its prompt and responses stay hidden. Our implementation infers the setting, held in a cached system-prompt prefix, from a co-resident process's own GPU kernel timing. On an eager sparse-MoE runtime, a frozen coverage study identified the deployed rule at \PcBasicBA balanced accuracy for its primary estimand and at \BhDecBA on tasks whose decision is the same under both rules, against \BhDecRestBA on tasks where the decisions differ. Routing replays and explicit transport locate the information in the temporal density and layer-and-bucket alignment of expert routing.

The disclosure reveals which deployments instruct the model to proceed without human confirmation and can inform target selection for subsequent jailbreak or prompt-injection attacks. For operators, this makes policy confidentiality a property to measure alongside behavioral compliance. For runtime developers, the dependence of execution on the active rule places policy-state distinguishability next to serving cost as an evaluation criterion. Our policy-placement comparison suppressed the evaluated timing classifier by moving the rule to a broker with trusted operation and authorization metadata (Section~\ref{sec:externalguard}). Applying that design to open-ended agents requires the framework to establish those facts about proposed operations; permission to access a resource alone does not establish that its use serves the authorized purpose.

These results cover one sparse-MoE model and known candidate rules profiled on the exact task, with request order and grouping available to the attacker.

\section*{Ethics Statement}\label{sec:ethics}

This work uses synthetic tasks and public models, with no human subjects or personal data. Hardware measurements run on research testbeds. The hosted-model behavior screen executes tool calls in disposable containers without host access. The native action interface records dry-run decisions, so the described deletions, external transmissions, and production changes are not executed. The attack requires co-residence on the agent's GPU instance, exact-task profiling, and invasive probing, and we report these conditions, the observation cost, and the evaluated padding tradeoff alongside the disclosure result so that operators can assess their exposure. All measurements use our own deployments on research testbeds.

\bibliographystyle{IEEEtran}
\bibliography{refs}

\end{document}